\documentclass[onecolumn,nofootinbib,amsmath,amssymb,aps,prd]{revtex4-2}
\usepackage{hyperref}
\begin{document}
\title{A principal-symbol obstruction in metric-affine Gauss--Bonnet gravity}
\author{Hamed Bouzari Nezhad}
\altaffiliation{Independent researcher}
\email{hamed.bouzarinezhad@gmail.com}
\affiliation{Brussels, Belgium}
\date{\today}
\begin{abstract}
The four-dimensional Riemannian Gauss--Bonnet combination admits more than one continuation to a geometry with an independent affine connection. Two natural candidates, the metric-contracted representative and the double-epsilon representative, agree in their Levi-Civita limit, and we ask whether they also give the independent connection the same leading differential structure. In a metric-affine model with a scalar coupling $f(\phi)$ to the Gauss--Bonnet term, we compare their quadratic two-derivative connection principal forms about the Levi-Civita locus. For the metric-contracted continuation, the principal form, evaluated on a twenty-four-dimensional hook-free torsion-free witness subspace at a timelike principal covector, has rank $18$ and nullity $6$ and is indefinite; the full torsion-free form at the same covector is therefore indefinite with rank at least $18$. Locally where $f(\phi)\neq0$, the two-derivative principal symbol of the linearized connection equations is nonzero, which obstructs purely algebraic local elimination of the connection in the activated directions. For the double-epsilon continuation, the corresponding quadratic two-derivative form vanishes identically for an arbitrary perturbation of the connection about that locus, so its principal symbol vanishes with it, and that particular obstruction is absent, though its absence does not by itself establish algebraic eliminability. Thus the leading differential structure of the independent connection depends on the affine continuation of the Riemannian Gauss--Bonnet combination.
\end{abstract}
\maketitle
\section{Introduction}
\label{Sec: Introduction}
In four dimensions the Gauss--Bonnet combination $\mathring{\mathcal{G}}=\mathring{R}^{2}-4\mathring{R}_{\alpha\beta}\mathring{R}^{\alpha\beta}+\mathring{R}_{\alpha\beta\gamma\delta}\mathring{R}^{\alpha\beta\gamma\delta}$,\footnote{Throughout this paper, a ring marks a Levi-Civita quantity, and the corresponding unringed quantities are those of the independent connection.} built from the Levi-Civita curvature of the metric, is the four-dimensional Euler density, and the scalar density $\sqrt{-g}\mathring{\mathcal{G}}$ is locally a total derivative \cite{Lanczos:1938sf,Lovelock:1971yv,Deruelle:2018vtt}, whereas $\sqrt{-g}f(\phi)\mathring{\mathcal{G}}$ is not. In the metric formulation the latter coupling belongs to the Horndeski class \cite{Horndeski:1974wa,Kobayashi:2011nu}. For what follows, the relevant point is that the Levi-Civita curvature satisfies antisymmetry within each index pair, $\mathring{R}_{\beta\alpha\gamma\delta}=-\mathring{R}_{\alpha\beta\gamma\delta}$ and $\mathring{R}_{\alpha\beta\delta\gamma}=-\mathring{R}_{\alpha\beta\gamma\delta}$; exchange of the two pairs, $\mathring{R}_{\alpha\beta\gamma\delta}=\mathring{R}_{\gamma\delta\alpha\beta}$; and the first Bianchi identity, $\mathring{R}_{[\alpha\beta\gamma]\delta}=0$.

For an independent connection $\Gamma^{\gamma}{}_{\alpha\beta}$ these symmetries are not all available. We use the curvature convention
\begin{eqnarray}
\label{Eq: Curvature convention}
R_{\alpha\beta\gamma}{}^{\delta}(\Gamma)=\partial_{\beta}\Gamma^{\delta}{}_{\alpha\gamma}-\partial_{\alpha}\Gamma^{\delta}{}_{\beta\gamma}+\Gamma^{\delta}{}_{\beta\epsilon}\Gamma^{\epsilon}{}_{\alpha\gamma}-\Gamma^{\delta}{}_{\alpha\epsilon}\Gamma^{\epsilon}{}_{\beta\gamma},
\end{eqnarray}
with Ricci tensor $R_{\alpha\beta}=R_{\alpha\gamma\beta}{}^{\gamma}$ and Ricci scalar $R=g^{\alpha\beta}R_{\alpha\beta}$. With this index ordering the antisymmetric pair is the first one, $R_{\beta\alpha\gamma}{}^{\delta}=-R_{\alpha\beta\gamma}{}^{\delta}$, and the first Bianchi identity survives whenever the torsion vanishes. Once the upper index is lowered, however, antisymmetry in the second pair and exchange of the two pairs are both lost: in general $R_{\alpha\beta\delta\gamma}\neq-R_{\alpha\beta\gamma\delta}$ and $R_{\gamma\delta\alpha\beta}\neq R_{\alpha\beta\gamma\delta}$. The separation studied in this paper originates in the loss of these two symmetries and not in any failure of the first Bianchi identity.

These are exactly the symmetries that make the different contraction patterns in $\mathring{\mathcal{G}}$ coincide. Once these symmetries are lost, the Riemannian combination has no unique continuation to an independent connection \cite{Borunda:2008kf,BeltranJimenez:2014iie}. Two natural candidates are
\begin{eqnarray}
\label{Eq: Metric-contracted Gauss-Bonnet scalar}
\mathcal{G}_{\rm{met}}(\Gamma)&=&R^{2}-4R_{\alpha\beta}R^{\alpha\beta}+R_{\alpha\beta\gamma\delta}R^{\alpha\beta\gamma\delta},\\
\label{Eq: Euler-Pfaffian density}
\mathcal{E}_{\rm{Euler}}(\Gamma)&=&-\frac{1}{4}\epsilon^{\alpha\beta\mu\nu}\epsilon^{\gamma\delta\rho\sigma}R_{\alpha\beta\gamma\delta}R_{\mu\nu\rho\sigma},
\end{eqnarray}
where $\epsilon_{\alpha\beta\gamma\delta}$ is the Levi-Civita tensor of the metric, with $\epsilon_{0123}=+\sqrt{-g}$ and $\epsilon^{0123}=-1/\sqrt{-g}$. The subscript in $\mathcal{G}_{\rm{met}}$ denotes the contraction pattern, not the formulation: the curvatures are those of the independent connection, and their indices are contracted with the metric. We call \eqref{Eq: Metric-contracted Gauss-Bonnet scalar} the metric-contracted representative and \eqref{Eq: Euler-Pfaffian density} the double-epsilon representative. On a Levi-Civita connection the two expressions agree and both reduce to $\mathring{\mathcal{G}}$. Under nonmetricity they are no longer forced to agree, because the double-epsilon contraction antisymmetrizes both index pairs of each curvature while the metric contraction does not.

This paper compares the two continuations away from the Levi-Civita locus, through the connection field equation, and asks whether that equation has the same leading differential structure in the two cases. What is at stake is the purely algebraic local elimination of the connection: solving the connection equations pointwise and substituting the result back, which would reduce the model locally to a metric scalar--tensor theory \cite{Vitagliano:2010sr}. The relevant object is the two-derivative part of the connection equations linearized about the Levi-Civita locus: the connection principal symbol. Where this symbol is nonzero, the connection equations are differential at leading order in the directions it activates, and those directions are not eliminable in that way.

Inequivalence between the metric and Palatini formulations of scalar fields coupled to curvature is already known from explicit models \cite{Bauer:2008zj,Rasanen:2017ivk,Rasanen:2018ihz,Nezhad:2023dys}. The ambiguity considered here sits one step earlier, in the choice of the affine object itself. Already at the level of the action, the affine Gauss--Bonnet term is generally not a total derivative once nonmetricity is present \cite{Janssen:2019uao,JimenezCano:2021rlu,Iosifidis:2021crj}. In the Palatini formulation a scalar-coupled Gauss--Bonnet term has been studied through cosmological solutions and order reduction \cite{Hassan:2026meg}, while the linearized spectra of general metric-affine quadratic-curvature actions have been analyzed for ghosts, tachyons, and instabilities \cite{BeltranJimenez:2019acz,BeltranJimenez:2020sqf,Percacci:2020ddy}. Those analyses start from a given action, whereas the point here is which action the Gauss--Bonnet coupling defines in the first place.

We expand the connection about its Levi-Civita value and compute, for each representative, the quadratic form in the distortion that carries two derivatives. For the metric-contracted representative this is done on the hook-free torsion-free nonmetricity sector, which is a linear witness subspace, and for a timelike principal covector. On that subspace the principal form is already nonzero and indefinite once the nonvanishing coefficient $f(\phi)$ is factored out, which is enough to establish indefiniteness of the full torsion-free principal form. Where $f(\phi)\neq0$, the connection is therefore not eliminable by purely algebraic local relations in the directions this form activates. For the double-epsilon representative the corresponding quadratic two-derivative form vanishes identically, for an arbitrary perturbation of the connection about its Levi-Civita value, so its principal symbol vanishes with it; the vanishing does not by itself establish algebraic eliminability. The two continuations of $\mathring{\mathcal{G}}$ thus differ already in the leading derivative structure of the connection sector. These statements concern that structure and do not constitute a count of propagating modes.

The paper is organized as follows. Section \ref{Sec: Connection principal symbol} sets up the connection principal symbol and the elimination diagnostic. Section \ref{Sec: Hook-free nonmetricity sector} introduces the hook-free witness sector and derives the covariant principal form on it, which Sec. \ref{Sec: Timelike block decomposition} evaluates in little-group blocks for a timelike principal covector; the same diagnostic is applied to the double-epsilon representative in Sec. \ref{Sec: Double-epsilon representative}, and the directions left null by the metric-contracted form are examined in Sec. \ref{Sec: Principal-null directions}. Section \ref{Sec: Discussion and conclusions} interprets the comparison, and Appendix \ref{AppSec: Component evaluation of the quadratic forms} records the component checks.
\section{Connection principal symbol}
\label{Sec: Connection principal symbol}
\subsection{Action and assumptions}
\label{Subsec: Action and assumptions}
We work in four dimensions with signature $(-,+,+,+)$ and take the affine connection $\Gamma^{\gamma}{}_{\alpha\beta}$ to be independent of the metric \cite{Hehl:1994ue}. The action is
\begin{eqnarray}
\label{Eq: Action}
S=\int d^{4}x\sqrt{-g}\left[\frac{M^{2}}{2}R(\Gamma)-\frac{1}{2}(\partial\phi)^{2}-V(\phi)+f(\phi)\mathcal{G}_{\rm{met}}(\Gamma)\right],
\end{eqnarray}
with $M^{2}>0$ and $\mathcal{G}_{\rm{met}}$ the metric-contracted representative \eqref{Eq: Metric-contracted Gauss-Bonnet scalar}, whose curvatures are those of $\Gamma^{\gamma}{}_{\alpha\beta}$. Throughout this paper, we restrict to a torsion-free connection, so that the departure of the independent connection from its Levi-Civita value is carried entirely by the nonmetricity. The same construction with \eqref{Eq: Euler-Pfaffian density} in its place is treated in Sec. \ref{Sec: Double-epsilon representative}.
\subsection{Distortion expansion and principal symbol}
\label{Subsec: Distortion expansion and principal symbol}
We decompose the independent connection into its Levi-Civita part and the distortion,
\begin{eqnarray}
\label{Eq: Distortion expansion}
\Gamma^{\gamma}{}_{\alpha\beta}=\mathring{\Gamma}^{\gamma}{}_{\alpha\beta}+L^{\gamma}{}_{\alpha\beta},
\end{eqnarray}
and expand the action \eqref{Eq: Action} to quadratic order in the distortion about the Levi-Civita locus $L^{\gamma}{}_{\alpha\beta}=0$. For a torsion-free connection the distortion is fixed by the nonmetricity $Q_{\gamma\alpha\beta}=\nabla_{\gamma}g_{\alpha\beta}$ through
\begin{eqnarray}
\label{Eq: Distortion from nonmetricity}
L_{\alpha\beta\gamma}=\frac{1}{2}\left(Q_{\alpha\beta\gamma}-Q_{\beta\alpha\gamma}-Q_{\gamma\alpha\beta}\right),
\end{eqnarray}
where $L_{\alpha\beta\gamma}=g_{\alpha\delta}L^{\delta}{}_{\beta\gamma}$, which is symmetric in its last two indices. Substituting the decomposition \eqref{Eq: Distortion expansion} into the curvature convention \eqref{Eq: Curvature convention} gives
\begin{eqnarray}
\label{Eq: Curvature expansion}
R_{\alpha\beta\gamma}{}^{\delta}=\mathring{R}_{\alpha\beta\gamma}{}^{\delta}-2\mathring{\nabla}_{[\alpha}L^{\delta}{}_{\beta]\gamma}-2L^{\delta}{}_{[\alpha}{}^{\epsilon}L_{|\epsilon|\beta]\gamma}.
\end{eqnarray}

The term linear in $L^{\gamma}{}_{\alpha\beta}$ carries one derivative of the distortion, and the term quadratic in it carries none. Inserting two copies of the linear term into the curvature-squared interaction therefore produces a quadratic form of type $(\mathring{\nabla}L)^{2}$, which is the only source of two derivatives of the distortion at this order. The Einstein--Hilbert sector supplies none: its term linear in $\mathring{\nabla}_{\delta}L^{\gamma}{}_{\alpha\beta}$ is a total derivative, and the quadratic distortion terms that remain contain no derivatives of $L^{\gamma}{}_{\alpha\beta}$ \cite{Shimada:2018lnm}. The two-derivative quadratic form therefore comes from the Gauss--Bonnet sector alone, with coefficient $f(\phi)$, and is present locally where $f(\phi)\neq0$. To obtain the corresponding linearized connection equations, we vary this quadratic form with respect to the distortion and integrate by parts. A derivative can then act on the coefficient $f(\phi)$ rather than on the distortion; these contributions, schematically $(\partial f)\mathring{\nabla}L$, carry only one derivative of the distortion and so do not contribute to the two-derivative principal symbol. Where the scalar background makes that gradient nonzero, these lower-order terms can enter the equations along the principal-null directions examined in Sec. \ref{Sec: Principal-null directions}.

The leading derivative structure is extracted in the usual way. Background coefficients are frozen at the point considered; only the terms of highest order in derivatives of the perturbation are retained, and each derivative acting on the perturbation is replaced by a covector, $\mathring{\nabla}_{\alpha}\to k_{\alpha}$ \cite{Reall:2014pwa,Papallo:2017qvl}; terms of lower derivative order drop out by construction. A hat marks a quantity obtained in this way. Applied to the linear term of \eqref{Eq: Curvature expansion}, this gives the principal curvature
\begin{eqnarray}
\label{Eq: Principal curvature}
\widehat{R}_{\alpha\beta\gamma}{}^{\delta}=-k_{\alpha}L^{\delta}{}_{\beta\gamma}+k_{\beta}L^{\delta}{}_{\alpha\gamma}.
\end{eqnarray}

This changes sign under $\alpha\leftrightarrow\beta$, as \eqref{Eq: Curvature convention} requires. Lowering the last index does not produce antisymmetry in the second pair: for the Levi-Civita curvature that antisymmetry follows from metric compatibility, and once nonmetricity is present, nothing enforces it, so $\widehat{R}_{\alpha\beta\gamma\delta}$ and $-\widehat{R}_{\alpha\beta\delta\gamma}$ need not agree.

Substituting the principal curvature \eqref{Eq: Principal curvature} into the metric-contracted combination \eqref{Eq: Metric-contracted Gauss-Bonnet scalar} gives the quadratic principal form
\begin{eqnarray}
\label{Eq: Principal symbol of the metric-contracted scalar}
\widehat{\mathcal{G}}_{\rm{met}}=\widehat{R}^{2}-4\widehat{R}_{\alpha\beta}\widehat{R}^{\alpha\beta}+\widehat{R}_{\alpha\beta\gamma\delta}\widehat{R}^{\alpha\beta\gamma\delta},
\end{eqnarray}
with $\widehat{R}_{\alpha\beta}=\widehat{R}_{\alpha\gamma\beta}{}^{\gamma}$ and $\widehat{R}=g^{\alpha\beta}\widehat{R}_{\alpha\beta}$. For the block analysis below, $\widehat{\mathcal{G}}_{\rm{met}}$ denotes the coefficient-free quadratic principal form, while its contribution to the action is $f(\phi)\widehat{\mathcal{G}}_{\rm{met}}$. In a basis $\mathbf{x}$ of distortion components we write $\widehat{\mathcal{G}}_{\rm{met}}=\mathbf{x}^{T}M(k)\mathbf{x}$, with $M(k)$ the symmetric matrix representing this form. At fixed metric, \eqref{Eq: Distortion expansion} gives $\delta\Gamma^{\gamma}{}_{\alpha\beta}=\delta L^{\gamma}{}_{\alpha\beta}$, so the connection variation may equivalently be taken with respect to the distortion. That variation is taken within the torsion-free connection space of Sec. \ref{Subsec: Action and assumptions}, so $\delta L^{\gamma}{}_{\alpha\beta}$ is symmetric in its last two indices and the connection equations are those of the torsion-free model. Since $M(k)$ is symmetric, varying the principal quadratic form gives $\delta\widehat{\mathcal{G}}_{\rm{met}}=2\delta\mathbf{x}^{T}M(k)\mathbf{x}$. The two-derivative principal symbol of the linearized connection equations is therefore proportional to $f(\phi)M(k)$. Wherever $f(\phi)\neq0$, this principal symbol has the same rank, nullity, and null directions as $M(k)$.

Where that matrix is nonzero, the linearized equations for the directions it activates contain derivatives of the connection at leading order rather than pointwise algebraic relations among the fields, so those components cannot be solved for and substituted back by a purely algebraic local procedure. The implication runs one way: a vanishing two-derivative symbol removes only this obstruction and does not by itself establish algebraic eliminability. The conclusion here concerns only direct local algebraic elimination of the connection.

To evaluate this diagnostic explicitly, one would in principle work with the forty independent components of a torsion-free distortion. For the present purpose, Sec. \ref{Sec: Hook-free nonmetricity sector} restricts the principal form to the hook-free linear subspace, which is used there as a witness subspace.
\section{Hook-free nonmetricity sector}
\label{Sec: Hook-free nonmetricity sector}
\subsection{Hook-free nonmetricity decomposition}
\label{Subsec: Hook-free nonmetricity decomposition}
The nonmetricity $Q_{\alpha\beta\gamma}=\nabla_{\alpha}g_{\beta\gamma}$ is symmetric in its last two indices because the metric is symmetric, independently of any condition on the torsion. In the torsion-free sector studied here it supplies the complete parametrization of the distortion through \eqref{Eq: Distortion from nonmetricity}, and its forty components carry two independent traces,
\begin{eqnarray}
\label{Eq: Nonmetricity traces}
Q_{\alpha}=g^{\beta\gamma}Q_{\alpha\beta\gamma},\qquad
\widetilde{Q}_{\alpha}=g^{\beta\gamma}Q_{\beta\gamma\alpha}.
\end{eqnarray}

Its irreducible content is a totally symmetric part, a mixed-symmetry (hook) part, and the two trace vectors \cite{McCrea:1992wa,Hehl:1994ue,Paci:2023twc,Mikura:2024mji}. We restrict to the \emph{hook-free} sector, setting the hook part to zero and keeping
\begin{eqnarray}
\label{Eq: Hook-free decomposition}
Q_{\alpha\beta\gamma}=\Omega_{\alpha\beta\gamma}+a_{\alpha}g_{\beta\gamma}+b_{(\beta}g_{|\alpha|\gamma)},\qquad
\Omega_{\alpha\beta\gamma}=\Omega_{(\alpha\beta\gamma)},\qquad
g^{\beta\gamma}\Omega_{\alpha\beta\gamma}=0,
\end{eqnarray}
where $\Omega_{\alpha\beta\gamma}$ is the totally symmetric traceless part, while $a_{\alpha}$ and $b_{\alpha}$ are the vector coefficients carrying the trace information. They are not independent of the traces: taking the two contractions \eqref{Eq: Nonmetricity traces} of \eqref{Eq: Hook-free decomposition} fixes them as
\begin{eqnarray}
\label{Eq: Trace coefficients}
a_{\alpha}=\frac{1}{18}(5Q_{\alpha}-2\widetilde{Q}_{\alpha}),\qquad
b_{\alpha}=\frac19(4\widetilde{Q}_{\alpha}-Q_{\alpha}).
\end{eqnarray}

A totally symmetric rank-three tensor in four dimensions has twenty independent components, and its trace is a four-vector, so tracelessness leaves sixteen independent components in $\Omega_{\alpha\beta\gamma}$. Including the two four-component trace vectors, the hook-free sector therefore has twenty-four independent components:
\begin{eqnarray}
\label{Eq: Hook-free sector dimension}
16_{\Omega}+4_{Q}+4_{\widetilde{Q}}=24.
\end{eqnarray}

The discarded hook part carries the remaining sixteen components, so \eqref{Eq: Hook-free decomposition} selects a twenty-four-dimensional linear subspace of the forty-dimensional torsion-free perturbation space.

The hook-free sector is used only as a witness subspace of the full torsion-free connection space. We evaluate the principal form explicitly on this subspace because its nonzero and indefinite character there is already enough to establish the corresponding properties of the full form. We do not assume that the hook-free sector is dynamically closed or defines a consistent truncation of the full theory. Let $M(k)$ be the symmetric matrix of the coefficient-free principal form on the full torsion-free space and $M(k)|_{\rm{hf}}$ its restriction to the hook-free subspace. Three properties of the full form follow directly from the restriction. If the restriction is nonzero, the full form is nonzero, since every vector in the hook-free subspace is also a vector of the full space. If the restriction is indefinite, the full form is indefinite, since two subspace vectors on which it takes opposite signs retain those values in the full space. Likewise,
\begin{eqnarray}
\label{Eq: Rank lower bound}
\operatorname{rank}M(k)\geq\operatorname{rank}M(k)|_{\rm{hf}},
\end{eqnarray}
because, in a basis adapted to the hook-free subspace, the restricted matrix is a principal submatrix of $M(k)$ and therefore cannot have larger rank.

Null directions do not transfer in the same way. In a basis in which the hook-free variables precede the hook variables, the full matrix takes the block form
\begin{eqnarray}
\label{Eq: Hook-free block decomposition}
M(k)=
\begin{pmatrix}
A & B\\
B^{T} & C
\end{pmatrix},
\qquad
A=M(k)|_{\rm{hf}},
\end{eqnarray}
in which a hook-free vector has the form $(v,0)$. For such a vector with $Av=0$,
\begin{eqnarray}
\label{Eq: Full action on restricted null vector}
M(k)(v,0)^{T}=(0,B^{T}v)^{T},
\end{eqnarray}
which need not vanish: the off-diagonal block $B$ couples $v$ to the hook directions set to zero in the restriction. A restricted principal-null direction therefore need not be null for the full form, and the nullity of the hook-free restriction does not determine the nullity of the full torsion-free form.

The trace variables can be organized with the help of the projective shift $\delta_{\xi}\Gamma^{\gamma}{}_{\alpha\beta}=\delta_{\beta}{}^{\gamma}\xi_{\alpha}$ \cite{Hehl:1994ue,Janssen:2019doc}, under which
\begin{eqnarray}
\label{Eq: Projective variations}
\delta_{\xi}Q_{\alpha\beta\gamma}=-2\xi_{\alpha}g_{\beta\gamma},\qquad
\delta_{\xi}Q_{\alpha}=-8\xi_{\alpha},\qquad
\delta_{\xi}\widetilde{Q}_{\alpha}=-2\xi_{\alpha},\qquad
\delta_{\xi}\Omega_{\alpha\beta\gamma}=0.
\end{eqnarray}

The combination
\begin{eqnarray}
\label{Eq: Projectively invariant trace}
P_{\alpha}=Q_{\alpha}-4\widetilde{Q}_{\alpha},\qquad
\delta_{\xi}P_{\alpha}=0,
\end{eqnarray}
is invariant under this shift, so $(P_{\alpha},\widetilde{Q}_{\alpha})$ provides a convenient projective-adapted trace basis. With this choice, $\delta_{\xi}a_{\alpha}=-2\xi_{\alpha}$ and $\delta_{\xi}b_{\alpha}=0$.

The projective shift is used only to select this trace basis. It does not preserve the torsion-free sector, since $\delta_{\xi}T^{\gamma}{}_{\alpha\beta}=\delta_{\beta}{}^{\gamma}\xi_{\alpha}-\delta_{\alpha}{}^{\gamma}\xi_{\beta}$ is generically nonzero, and $\mathcal{G}_{\rm{met}}$ is not projectively invariant in general. It is therefore not a symmetry of the torsion-free theory considered here, and connections related by this shift are not identified as equivalent. So the invariance of $P_{\alpha}$ under the shift does not make it a gauge-invariant physical mode.
\subsection{Covariant principal form}
\label{Subsec: Covariant principal form}
Substituting \eqref{Eq: Hook-free decomposition} and \eqref{Eq: Trace coefficients} into \eqref{Eq: Distortion from nonmetricity} expresses the distortion through the hook-free variables,
\begin{eqnarray}
\label{Eq: Reduced distortion}
L_{\alpha\beta\gamma}=-\frac{1}{2}\Omega_{\alpha\beta\gamma}+\frac{1}{2}c_{\alpha}g_{\beta\gamma}-a_{(\beta}g_{|\alpha|\gamma)},
\end{eqnarray}
with $c_{\alpha}=a_{\alpha}-b_{\alpha}$. In the projective-adapted basis $a_{\alpha}=\left(5P_{\alpha}/18\right)+\widetilde{Q}_{\alpha}$, $b_{\alpha}=-P_{\alpha}/9$, and $c_{\alpha}=\left(7P_{\alpha}/18\right)+\widetilde{Q}_{\alpha}$, so $L_{\alpha\beta\gamma}$ depends only on $\Omega_{\alpha\beta\gamma}$, $P_{\alpha}$, and $\widetilde{Q}_{\alpha}$.

Two contractions of $\Omega_{\alpha\beta\gamma}$ with the principal covector appear, $\Pi_{\alpha\beta}=k^{\gamma}\Omega_{\alpha\beta\gamma}$ and $\Sigma_{\alpha}=k^{\beta}k^{\gamma}\Omega_{\alpha\beta\gamma}$. We write $|\Omega|^{2}=\Omega_{\alpha\beta\gamma}\Omega^{\alpha\beta\gamma}$ and $x\cdot y=x_{\alpha}y^{\alpha}$ with $x^{2}=x\cdot x$ for scalar products of vectors, so that $k^{2}=k\cdot k$. Substituting \eqref{Eq: Reduced distortion} into the principal curvature \eqref{Eq: Principal curvature} and the result into \eqref{Eq: Principal symbol of the metric-contracted scalar} gives the exact hook-free principal form
\begin{eqnarray}
\label{Eq: Exact hook-free principal quadratic form}
\widehat{\mathcal{G}}_{\rm{met}}&=&\frac{1}{2}k^{2}|\Omega|^{2}-\frac{3}{2}\Pi_{\alpha\beta}\Pi^{\alpha\beta}+6(k\cdot\widetilde{Q})^{2}-6k^{2}\widetilde{Q}^{2}+\Sigma_{\alpha}(4\widetilde{Q}^{\alpha}+P^{\alpha})\nonumber\\&&
+\frac{29}{9}(k\cdot\widetilde{Q})(k\cdot P)-\frac{23}{9}k^{2}(\widetilde{Q}\cdot P)+\frac{47}{108}(k\cdot P)^{2}-\frac{13}{54}k^{2}P^{2}.
\end{eqnarray}

Setting $P_{\alpha}=\widetilde{Q}_{\alpha}=0$ isolates the totally symmetric traceless component and leaves
\begin{eqnarray}
\label{Eq: Principal symbol in the symmetric sector}
\widehat{\mathcal{G}}_{\rm{met}}\big|_{\Omega}=\frac{1}{2}k^{2}|\Omega|^{2}-\frac{3}{2}\Pi_{\alpha\beta}\Pi^{\alpha\beta}.
\end{eqnarray}

This is the part of the principal form that does not involve the trace vectors, displayed separately before their contributions are included. The full covariant hook-free form \eqref{Eq: Exact hook-free principal quadratic form} holds for an arbitrary principal covector; Sec. \ref{Sec: Timelike block decomposition} specializes it to a timelike covector and decomposes it into little-group blocks.
\section{Timelike block decomposition}
\label{Sec: Timelike block decomposition}
We now take the principal covector timelike and work in its rest frame, $k_{\alpha}=(\omega,0,0,0)$, so that $k^{2}=-\omega^{2}$. The little group of the timelike covector $k_{\alpha}$ is the rotation group $SO(3)$. Spatial indices in this frame are denoted by $i,j,k,\ldots=1,2,3$. The variables are then decomposed into irreducible representations of $SO(3)$, labeled by spin $J$, with dimension $2J+1$ \cite{Percacci:2020ddy,Baldazzi:2021kaf}. All rank, nullity, and inertia statements in this section are established for timelike principal covectors only.

Under $SO(3)$ the totally symmetric traceless $\Omega_{\alpha\beta\gamma}$ splits as
\begin{eqnarray}
\label{Eq: Timelike decomposition of Omega}
&&\Omega_{ijk}=T_{ijk}+\frac{1}{5}\left(\delta_{ij}v_{k}+\delta_{ik}v_{j}+\delta_{jk}v_{i}\right),\nonumber\\&&
\Omega_{0ij}=S_{ij}+\frac{1}{3}\delta_{ij}s,\qquad
\Omega_{00i}=v_{i},\qquad
\Omega_{000}=s,
\end{eqnarray}
where $T_{ijk}=T_{(ijk)}$ is a spatially traceless symmetric rank-three tensor, $S_{ij}=S_{(ij)}$ a spatially traceless symmetric rank-two tensor, $v_{i}$ a spatial vector, and $s$ a scalar. These carry seven, five, three, and one components, respectively, so the sixteen components of $\Omega_{\alpha\beta\gamma}$ split as $16=7+5+3+1$ with spin $3,2,1,0$. Each trace vector splits into a spatial part and a time component, $P_{\alpha}\to(P_{i},P_{0})$ and $\widetilde{Q}_{\alpha}\to(\widetilde{Q}_{i},\widetilde{Q}_{0})$, contributing spin $1$ and spin $0$. The twenty-four witness variables therefore carry spin $3$ in the seven components of $T_{ijk}$, spin $2$ in the five of $S_{ij}$, spin $1$ in the nine of $v_{i}$, $P_{i}$, and $\widetilde{Q}_{i}$, and spin $0$ in the three of $s$, $P_{0}$, and $\widetilde{Q}_{0}$:
\begin{eqnarray}
\label{Eq: Hook-free little-group content}
7_{J=3}+5_{J=2}+9_{J=1}+3_{J=0}=24.
\end{eqnarray}

For the timelike principal covector $k_{\alpha}=(\omega,0,0,0)$, the $SO(3)$ little group leaves the background metric and $k_{\alpha}$ unchanged while rotating the distortion components among themselves as spatial tensors. Since $\widehat{\mathcal{G}}_{\rm{met}}$ is a scalar built covariantly from the metric, $k_{\alpha}$, and the distortion, its value is unchanged under this induced action on the distortion. This allows the distortion to be decomposed into irreducible spin-$J$ sectors, and an invariant quadratic form cannot couple inequivalent representations, so sectors of different spin do not mix. A given spin can nevertheless appear more than once, and copies of the same representation are free to mix among themselves. The vector sector shows this most clearly: $v_{i}$, $P_{i}$, and $\widetilde{Q}_{i}$ are three copies of the spin-1 representation, and invariance forces one and the same $3\times3$ matrix to act on the triple $(v_{i},P_{i},\widetilde{Q}_{i})$ at each spatial component $i$. Its nine-dimensional block is therefore $M^{(1)}\otimes I_{3}$, with $M^{(1)}$ mixing the three copies and $I_{3}$ acting on the spatial index, and one spatial direction already determines the whole block \cite{Percacci:2020ddy,Baldazzi:2021kaf}. The spin-0 sector is the three scalars $s$, $P_{0}$, and $\widetilde{Q}_{0}$, which carry no spatial index and give a single $3\times3$ matrix.

In each sector $M^{(J)}$ denotes the matrix of $\widehat{\mathcal{G}}^{(J)}_{\rm{met}}$ in the component basis stated for that sector, with the nonzero coefficient $f(\phi)$ factored out as in Sec. \ref{Sec: Connection principal symbol}. For $J=1$ and $J=0$ the matrices are written in the component bases stated below. The spin-3 sector contains a single copy of the seven-dimensional representation, so its contribution is written most naturally through the invariant norm $T_{ijk}T^{ijk}$ rather than as a $7\times7$ component matrix.
\subsection{Tensor sectors}
\label{Subsec: Tensor sectors}
On the spin-3 representation \cite{Baekler:2006vw,Percacci:2025oxw} the trace vectors are absent and every component of $\Omega_{\alpha\beta\gamma}$ carrying a time index vanishes, so $\Pi_{\alpha\beta}=k^{\gamma}\Omega_{\alpha\beta\gamma}=0$ and $|\Omega|^{2}=T_{ijk}T^{ijk}$. With $k^{2}=-\omega^{2}$, \eqref{Eq: Principal symbol in the symmetric sector} gives
\begin{eqnarray}
\label{Eq: Spin-3 principal form}
\widehat{\mathcal{G}}^{(3)}_{\rm{met}}=-\frac{1}{2}\omega^{2}T_{ijk}T^{ijk}.
\end{eqnarray}

The block is thus $-1/2$ times the invariant spin-3 norm. Since $T_{ijk}T^{ijk}$ is positive for every nonzero spin-3 tensor, the quadratic form is negative definite on this seven-dimensional sector and has rank seven, reducing to $M^{(3)}=-\frac{1}{2}I_{7}$ in a basis orthonormal with respect to that norm.

In the spin-2 sector the only nonzero irreducible variable is the symmetric traceless tensor $S_{ij}$, which appears in the components of $\Omega_{\alpha\beta\gamma}$ carrying a single time index. Total symmetry makes the three placements $\Omega_{0ij}$, $\Omega_{i0j}$, and $\Omega_{ij0}$ equal, and raising the time index in each contributes one factor $g^{00}=-1$, so
\begin{eqnarray}
\label{Eq: Spin-2 contractions}
|\Omega|^{2}=-3S_{ij}S^{ij},\qquad
\Pi_{\alpha\beta}\Pi^{\alpha\beta}=\omega^{2}S_{ij}S^{ij}.
\end{eqnarray}

After substituting these contractions into \eqref{Eq: Principal symbol in the symmetric sector} with $k^{2}=-\omega^{2}$, the two terms become $\frac{3}{2}\omega^{2}S_{ij}S^{ij}$ and $-\frac{3}{2}\omega^{2}S_{ij}S^{ij}$ and cancel exactly. The spin-2 block therefore vanishes, $M^{(2)}=0$ of rank zero, so the five components of $S_{ij}$ are restricted principal-null directions. Their Einstein--Hilbert structure is examined in Sec. \ref{Sec: Principal-null directions}.
\subsection{Vector and scalar sectors}
\label{Subsec: Vector and scalar sectors}
The spin-1 sector contains the three vector variables $v_{i}$, $P_{i}$, and $\widetilde{Q}_{i}$. For one fixed spatial direction $i$, with no sum on that index, and with the variables collected into $\mathbf{x}^{(1)}_{i}=(v_{i},P_{i},\widetilde{Q}_{i})^{T}$, the form becomes
\begin{eqnarray}
\label{Eq: Spin-1 block matrix}
\widehat{\mathcal{G}}^{(1)}_{\rm{met}}=\omega^{2}\mathbf{x}^{(1)T}_{i}M^{(1)}\mathbf{x}^{(1)}_{i},\qquad
M^{(1)}=
\begin{pmatrix}
\dfrac{6}{5} & \dfrac{1}{2} & 2\\[6pt]
\dfrac{1}{2} & \dfrac{13}{54} & \dfrac{23}{18}\\[6pt]
2 & \dfrac{23}{18} & 6
\end{pmatrix}.
\end{eqnarray}

With $\det{M^{(1)}}=-2/15$, $\operatorname{rank}M^{(1)}=3$, and inertia $(2_{+},1_{-})$, this block is nonsingular and indefinite, with two positive and one negative direction. The same block occurs for each of the three spatial directions, giving the vector sector rank $3\times3=9$.

The spin-0 sector contains the three scalars $s$, $P_{0}$, and $\widetilde{Q}_{0}$. Collecting them into $\mathbf{x}^{(0)}=(s,P_{0},\widetilde{Q}_{0})^{T}$ gives
\begin{eqnarray}
\label{Eq: Spin-0 block matrix}
\widehat{\mathcal{G}}^{(0)}_{\rm{met}}=\omega^{2}\mathbf{x}^{(0)T}M^{(0)}\mathbf{x}^{(0)},\qquad
M^{(0)}=
\begin{pmatrix}
-1 & -\dfrac{1}{2} & -2\\[6pt]
-\dfrac{1}{2} & \dfrac{7}{36} & \dfrac{1}{3}\\[6pt]
-2 & \dfrac{1}{3} & 0
\end{pmatrix}.
\end{eqnarray}

Unlike the vector block, this one is degenerate: $\det{M^{(0)}}=0$ and $\operatorname{rank}M^{(0)}=2$. Its nonzero eigenvalues $\lambda_{\neq0}=(-29\pm\sqrt{24457})/72$ have opposite signs, so the nonzero part has inertia $(1_{+},1_{-})$ and is again indefinite. The remaining one-dimensional kernel is spanned by
\begin{eqnarray}
\label{Eq: Scalar null vector}
n=\left(-\frac{1}{2},-3,1\right)^{T},
\end{eqnarray}
with $M^{(0)}n=0$ in the basis $\mathbf{x}^{(0)}$. Thus $n$ defines a single restricted principal-null scalar direction, mixing $s$ with both trace scalars $P_{0}$ and $\widetilde{Q}_{0}$. We return to this direction in Sec. \ref{Sec: Principal-null directions}, where its Einstein--Hilbert structure is examined. The explicit component reconstruction of the spin-1 and spin-0 matrices is given in Appendix \ref{AppSec: Component evaluation of the quadratic forms}.
\subsection{Rank, nullity, and interpretation}
\label{Subsec: Rank, nullity, and interpretation}
Combining the four spin sectors gives the rank. The spin-3 block contributes rank seven, the spin-2 block vanishes, the three spin-1 copies contribute rank nine, and the spin-0 block contributes rank two, so for a timelike principal covector
\begin{eqnarray}
\label{Eq: Principal rank}
\operatorname{rank}\!\left(M(k)|_{\rm{hf}}\right)=7_{J=3}+9_{J=1}+2_{J=0}=18.
\end{eqnarray}

The null directions come from the same blocks. The spin-2 sector contributes five, while the spin-0 sector contributes the single scalar direction $n$, giving
\begin{eqnarray}
\label{Eq: Principal nullity}
\operatorname{null}\!\left(M(k)|_{\rm{hf}}\right)=5_{J=2}+1_{J=0}=6.
\end{eqnarray}

These results refer to the twenty-four-dimensional hook-free restriction and to timelike principal covectors only. The nonzero part of the form is indefinite because the spin-1 and spin-0 blocks each contain both positive and negative directions.

The particular matrices used to display the blocks depend on the chosen component bases, but the rank, nullity, and inertia do not. Under a nonsingular change of variables, $M\mapsto C^{T}MC$, the numbers of positive, negative, and zero directions are unchanged. Individual matrix entries, eigenvalues, and the magnitude of the determinant need not be preserved.

The blocks above represent the coefficient-free form $\widehat{\mathcal{G}}_{\rm{met}}$. Wherever $f(\phi)\neq0$, restoring the factor $f(\phi)$ leaves the rank, nullity, and null directions unchanged. For $f(\phi)>0$ the positive and negative counts are those displayed above, while for $f(\phi)<0$ they are exchanged. The form is indefinite in either case.

Because the hook-free sector is a subspace of the full torsion-free space, the calculation also gives two conclusions about the full principal form. For the same timelike principal covector, the full form is indefinite, since the hook-free subspace already contains directions of both signs, and its rank is at least eighteen. The six null directions do not extend in the same way. They are null only within the hook-free restriction, and couplings to the omitted hook components can make them non-null in the full form. The full torsion-free nullity is therefore not determined by this calculation.

Finally, wherever $f(\phi)\neq0$, the linearized connection equations contain two-derivative terms in the directions on which the principal matrix acts nontrivially. By the diagnostic of Sec. \ref{Sec: Connection principal symbol}, those directions are therefore obstructed from purely algebraic local elimination of the connection. Section \ref{Sec: Double-epsilon representative} applies the same test to the double-epsilon representative.
\section{Double-epsilon representative}
\label{Sec: Double-epsilon representative}
We now apply the same principal-curvature construction to the double-epsilon representative \eqref{Eq: Euler-Pfaffian density}. Here the outcome can be established before any hook-free or timelike restriction is imposed.
\subsection{Vanishing principal form}
\label{Subsec: Vanishing principal form}
For the double-epsilon representative the quadratic two-derivative principal form vanishes identically. With the last index of the principal curvature \eqref{Eq: Principal curvature} lowered, $\widehat{R}_{\alpha\beta\gamma\delta}=2k_{[\beta}L_{|\delta|\alpha]\gamma}$, so the momentum factor in each curvature lies in its first index pair. Substituting two such curvatures into the double-epsilon contraction places both momentum factors in the first Levi-Civita tensor, and
\begin{eqnarray}
\label{Eq: Vanishing Euler principal symbol}
\widehat{\mathcal{E}}_{\rm{Euler}}&=&-\frac{1}{4}\epsilon^{\alpha\beta\mu\nu}\epsilon^{\gamma\delta\rho\sigma}\widehat{R}_{\alpha\beta\gamma\delta}\widehat{R}_{\mu\nu\rho\sigma}\nonumber\\
&=&-\epsilon^{\alpha\beta\mu\nu}\epsilon^{\gamma\delta\rho\sigma}k_{\beta}k_{\nu}L_{\delta\alpha\gamma}L_{\sigma\mu\rho}=0.
\end{eqnarray}

The last equality follows from
\begin{eqnarray}
\label{Eq: Momentum argument}
\epsilon^{\alpha\beta\mu\nu}k_{\beta}k_{\nu}=0,
\end{eqnarray}
since the Levi-Civita tensor is antisymmetric under $\beta\leftrightarrow\nu$, while $k_{\beta}k_{\nu}$ is symmetric.

This proof does not use the hook-free decomposition, a timelike restriction on $k_{\alpha}$, or the torsion-free symmetry of $L^{\gamma}{}_{\alpha\beta}$. It therefore holds for an arbitrary distortion perturbation in the quadratic expansion about the Levi-Civita locus $L^{\gamma}{}_{\alpha\beta}=0$. It does not establish the result for a background with nonzero distortion.

In the pure $\Omega$ sector, the metric-contracted representative has the nontrivial principal form \eqref{Eq: Principal symbol in the symmetric sector}. Evaluating the double-epsilon representative on the same sector therefore gives a direct comparison of how the two affine continuations treat these distortion components. Setting $P_{\alpha}=\widetilde{Q}_{\alpha}=0$ gives $L_{\alpha\beta\gamma}=-\Omega_{\alpha\beta\gamma}/2$, which is totally symmetric. Equation \eqref{Eq: Principal curvature} then makes $\widehat{R}_{\alpha\beta\gamma\delta}$ symmetric under $\gamma\leftrightarrow\delta$. The second Levi-Civita tensor is antisymmetric in that pair and therefore annihilates the curvature,
\begin{eqnarray}
\label{Eq: Last-pair-symmetry argument}
\epsilon^{\gamma\delta\rho\sigma}\widehat{R}_{\alpha\beta\gamma\delta}=0\quad\Longrightarrow\quad\widehat{\mathcal{E}}_{\rm{Euler}}\big|_{\Omega}=0.
\end{eqnarray}

This argument is specific to the pure $\Omega$ sector and is not needed for the general proof, but it makes the contrast transparent: the metric-contracted representative has a nontrivial principal form on this sector, while the double-epsilon principal form vanishes there.
\subsection{Consequences of the vanishing principal form}
\label{Subsec: Consequences of the vanishing principal form}
The preceding subsection showed that the double-epsilon quadratic two-derivative action form vanishes in the expansion about the Levi-Civita locus $L^{\gamma}{}_{\alpha\beta}=0$, and the two-derivative principal symbol of the linearized connection equations vanishes with it. A direct contraction over all twenty-four hook-free components gives the same vanishing coefficient by coefficient, as given in Appendix \ref{AppSec: Component evaluation of the quadratic forms}. This provides a component verification of the analytic argument rather than a separate proof.

The principal-symbol obstruction to purely algebraic local elimination identified in Sec. \ref{Sec: Connection principal symbol} is therefore absent for the double-epsilon representative about $L^{\gamma}{}_{\alpha\beta}=0$. This does not by itself establish that the connection can be eliminated algebraically, nor does it show that the affine double-epsilon term is topological or dynamically trivial. Lower-derivative terms are not determined by this two-derivative principal calculation, and the result does not extend to backgrounds with nonzero distortion.

At the same Levi-Civita locus, the two representatives differ at the level of this diagnostic. For a timelike principal covector, the metric-contracted principal form is already nonzero and indefinite on the hook-free witness subspace, so the obstruction to purely algebraic local elimination is present in the directions activated there. The double-epsilon principal form instead vanishes, so the same obstruction is absent.

The comparison leaves the restricted principal-null directions of the metric-contracted representative to be examined separately. In the hook-free timelike restriction, six directions are principal-null, but they need not remain null when the omitted hook components are restored. Section \ref{Sec: Principal-null directions} examines their lower-order Einstein--Hilbert structure.
\section{Principal-null directions}
\label{Sec: Principal-null directions}
The metric-contracted principal form leaves six directions null in the hook-free timelike restriction of Sec. \ref{Sec: Timelike block decomposition}: five in the spin-2 sector and one in the spin-0 sector. Within this restriction, the Gauss--Bonnet two-derivative principal form vanishes on these six directions, so the next question is what algebraic quadratic structure the Einstein--Hilbert term leaves there.

At quadratic order about $L^{\gamma}{}_{\alpha\beta}=0$ the Einstein--Hilbert term contributes no two-derivative principal form, as established in Sec. \ref{Sec: Connection principal symbol}: its derivative piece is a boundary term at this order, and the remaining quadratic contribution is algebraic in the distortion,
\begin{eqnarray}
\label{Eq: Einstein-Hilbert algebraic form}
\mathcal{B}_{\rm{EH}}=\frac{M^{2}}{2}g^{\alpha\beta}\left(L^{\gamma}{}_{\gamma\delta}L^{\delta}{}_{\alpha\beta}-L^{\gamma}{}_{\alpha\delta}L^{\delta}{}_{\gamma\beta}\right).
\end{eqnarray}

We now evaluate this Einstein--Hilbert algebraic form on the six directions left null by the hook-free timelike Gauss--Bonnet principal form. It is not the complete zero-derivative quadratic operator, because the Gauss--Bonnet sector also contributes zero-derivative terms. These terms and the one-derivative Gauss--Bonnet terms are discussed in Sec. \ref{Subsec: Lower-order terms on the principal-null directions}, where we ask how they can modify the equations along these six directions without changing the two-derivative principal symbol.
\subsection{Spin-2 null directions}
\label{Subsec: Spin-2 null directions}
The five spin-2 directions are the independent components of the spatially traceless symmetric $S_{ij}$, with $S_{33}=-S_{11}-S_{22}$. With these components collected as $\mathbf{x}^{(2)}=(S_{11},S_{22},S_{12},S_{13},S_{23})^{T}$ and with $\mathcal{B}^{(2)}_{\rm{EH}}=\mathbf{x}^{(2)T}H^{(2)}\mathbf{x}^{(2)}$, the matrix $H^{(2)}$ represents \eqref{Eq: Einstein-Hilbert algebraic form} on this five-dimensional subspace. Its eigenvalues are $M^{2}/2$ times
\begin{eqnarray}
\label{Eq: Einstein-Hilbert spin-2 eigenvalues}
\left\{\frac{9}{4},\frac{3}{2},\frac{3}{2},\frac{3}{2},\frac{3}{4}\right\},
\end{eqnarray}
all positive for $M^{2}>0$, so $H^{(2)}$ is nonsingular and positive definite; the matrix and its determinant are given in Appendix \ref{AppSec: Component evaluation of the quadratic forms}.

Thus, within the hook-free timelike restriction, the five spin-2 directions that are null in the Gauss--Bonnet two-derivative block have a nonsingular, positive Einstein--Hilbert algebraic form. This alone does not show that they are algebraic variables in the complete torsion-free equations: the remaining lower-order terms of Sec. \ref{Subsec: Lower-order terms on the principal-null directions}, including the zero-derivative Gauss--Bonnet contributions, are not evaluated here, and the omitted hook variables may couple to these directions at principal order.
\subsection{Spin-0 null direction}
\label{Subsec: Spin-0 null direction}
In the scalar basis $\mathbf{x}^{(0)}=(s,P_{0},\widetilde{Q}_{0})^{T}$ of Sec. \ref{Sec: Timelike block decomposition}, the Einstein--Hilbert algebraic form becomes
\begin{eqnarray}
\label{Eq: Einstein-Hilbert spin-0 form}
\mathcal{B}^{(0)}_{\rm{EH}}=\frac{M^{2}}{2}\left(\frac{11}{72}P_{0}^{2}+P_{0}\widetilde{Q}_{0}+\frac{3}{2}\widetilde{Q}_{0}^{2}+\frac{1}{2}s^{2}\right)=\mathbf{x}^{(0)T}H^{(0)}\mathbf{x}^{(0)},
\end{eqnarray}
with
\begin{eqnarray}
\label{Eq: Einstein-Hilbert spin-0 matrix}
H^{(0)}=\frac{M^{2}}{2}
\begin{pmatrix}
\dfrac{1}{2} & 0 & 0\\[6pt]
0 & \dfrac{11}{72} & \dfrac{1}{2}\\[6pt]
0 & \dfrac{1}{2} & \dfrac{3}{2}
\end{pmatrix}.
\end{eqnarray}

This matrix is nonsingular, with $\det{H^{(0)}}=-M^{6}/768\neq0$, so it has no kernel in the three-dimensional scalar space. Evaluation on the Gauss--Bonnet principal-null vector \eqref{Eq: Scalar null vector} of $M^{(0)}$ gives
\begin{eqnarray}
\label{Eq: Einstein-Hilbert action on the null direction}
n^{T}H^{(0)}n=0,\qquad
H^{(0)}n=\frac{M^{2}}{2}\left(-\frac{1}{4},\frac{1}{24},0\right)^{T}\neq0.
\end{eqnarray}

These two statements say different things. The first, $n^{T}H^{(0)}n=0$, is that the Einstein--Hilbert quadratic form vanishes on $n$ itself, so $n$ has no Einstein--Hilbert self-coupling. The second, $H^{(0)}n\neq0$, is that $n$ is not in the kernel of $H^{(0)}$, so it still couples algebraically to the other scalar directions.

To make this distinction explicit, we choose a scalar basis whose first direction is the principal-null vector $n$. Writing $\mathbf{x}^{(0)}=U\mathbf{y}$ with $\mathbf{y}=(y_{1},y_{2},y_{3})^{T}$, we have
\begin{eqnarray}
\label{Eq: Null-adapted scalar basis}
U=
\begin{pmatrix}
-\dfrac{1}{2} & 0 & 0\\[6pt]
-3 & 1 & 0\\[6pt]
1 & 0 & 1
\end{pmatrix},\qquad
U^{T}M^{(0)}U=
\begin{pmatrix}
0 & 0 & 0\\[6pt]
0 & \dfrac{7}{36} & \dfrac{1}{3}\\[6pt]
0 & \dfrac{1}{3} & 0
\end{pmatrix},\qquad
U^{T}H^{(0)}U=\frac{M^{2}}{2}
\begin{pmatrix}
0 & \dfrac{1}{24} & 0\\[6pt]
\dfrac{1}{24} & \dfrac{11}{72} & \dfrac{1}{2}\\[6pt]
0 & \dfrac{1}{2} & \dfrac{3}{2}
\end{pmatrix}.
\end{eqnarray}

The first column of $U$ is $n$, so $y_{1}$ is the coordinate along the Gauss--Bonnet principal-null scalar direction, while $y_{2}$ and $y_{3}$ complete the scalar basis. In $U^{T}M^{(0)}U$ the first row and column vanish, so the Gauss--Bonnet two-derivative form contains no term involving $y_{1}$. In $U^{T}H^{(0)}U$ the $(1,1)$ entry vanishes but the $(1,2)$ entry does not, so the Einstein--Hilbert algebraic form contains no $y_{1}^{2}$ self-term but does contain a cross-term in $y_{1}y_{2}$. In the terms evaluated here, $y_{1}$ therefore appears only linearly, and varying with respect to $y_{1}$ gives an algebraic relation involving $y_{2}$. This is why $y_{1}$ is Lagrange-multiplier-like for the two quadratic forms evaluated here. It does not establish that $y_{1}$ is a Lagrange multiplier of the complete theory.
\subsection{Lower-order terms on the principal-null directions}
\label{Subsec: Lower-order terms on the principal-null directions}
Within the hook-free timelike restriction, the Gauss--Bonnet two-derivative principal form vanishes on the six principal-null directions. Lower-derivative terms can therefore enter the equations along these directions even though they do not alter the two-derivative principal symbol.

Two kinds of lower-order Gauss--Bonnet term arise. The Gauss--Bonnet quadratic derivative term is schematically $f(\phi)(\mathring{\nabla}L)^{2}$. On variation, a derivative can act either on the distortion or on $f(\phi)$. Acting on the distortion gives the two-derivative operator $f\mathring{\nabla}^{2}L$, which is present wherever $f(\phi)\neq0$. Acting on the coupling gives $(\partial f)\mathring{\nabla}L$ when $\partial_{\alpha}f\neq0$, and this term carries one derivative of the distortion. The quadratic expansion of $f(\phi)\mathcal{G}_{\rm{met}}$ also contains zero-derivative terms, schematically $f\mathring{R}L^{2}$, in which the background Levi-Civita curvature is contracted with the part of \eqref{Eq: Curvature expansion} that is quadratic in the distortion and carries no derivative. Such terms can be present on a curved background, including when $f(\phi)$ is constant. Neither kind carries two derivatives of the distortion, so neither changes the rank, nullity, inertia, or null directions of the two-derivative principal symbol. Both can nevertheless contribute to the equations along the six directions on which the restricted two-derivative block vanishes. The zero-derivative Gauss--Bonnet terms in particular mean that $H^{(2)}$ and $H^{(0)}$ are the Einstein--Hilbert part of the zero-derivative quadratic operator and not the whole of it.

The spin decomposition of Sec. \ref{Sec: Timelike block decomposition} applies to the timelike two-derivative principal form. At lower order, the background gradient $\partial_{\alpha}f$ and the background Levi-Civita curvature provide additional tensors, so the spin-2 and spin-0 sectors need not remain decoupled. Until these terms are evaluated, we cannot conclude that the five spin-2 variables remain purely algebraic in the complete quadratic equations or that $y_{1}$ remains multiplier-like there.

There is also a separate issue associated with the hook-free restriction. The omitted hook variables may couple at principal order to the six directions that are null within the hook-free form. The present calculation does not determine those couplings.

Our calculation is a quadratic Lagrangian and principal-operator analysis about the Levi-Civita locus, not a Dirac--Hamiltonian constraint analysis. It therefore does not determine the Hamiltonian constraint structure, whether $y_{1}$ generates a genuine Hamiltonian constraint, or the number of propagating independent-connection degrees of freedom \cite{Glavan:2023cuy}.
\section{Discussion and conclusions}
\label{Sec: Discussion and conclusions}
The two affine continuations coincide on the Levi-Civita locus but differ in their quadratic two-derivative connection structure about it. For the metric-contracted representative, the principal form on the twenty-four-dimensional hook-free torsion-free witness subspace has rank $18$ and nullity $6$ and is indefinite for a timelike principal covector. The witness argument then gives, for the same covector, a full torsion-free form that is indefinite and has rank at least $18$; its nullity is not determined. Wherever $f(\phi)\neq0$, the two-derivative principal symbol of the linearized connection equations is nonzero in the directions this form activates, which obstructs purely algebraic local elimination of the connection there. For the double-epsilon representative, the coefficient-free quadratic two-derivative action form vanishes for an arbitrary distortion perturbation about $L^{\gamma}{}_{\alpha\beta}=0$, without the hook-free restriction, a timelike restriction on the principal covector, or torsion-free symmetry of the perturbation, and the equation-level principal symbol vanishes with it. The principal-symbol obstruction is therefore present for the metric-contracted representative and absent for the double-epsilon representative about the same locus, although its absence alone does not establish algebraic eliminability.

This diagnostic differs from the questions addressed in the metric-affine literature introduced earlier. Spectral and stability analyses of fixed metric-affine quadratic-curvature theories start from a chosen affine action \cite{BeltranJimenez:2019acz,BeltranJimenez:2020sqf,Percacci:2020ddy}. Here the question comes one step earlier: how the choice of affine continuation changes the connection principal structure. Studies of whether the affine Gauss--Bonnet term retains its total-derivative or topological character under nonmetricity concern the action-level structure rather than the connection principal symbol \cite{Janssen:2019uao,JimenezCano:2021rlu,Iosifidis:2021crj}. Nonlinear analyses of double-epsilon Palatini models provide a useful comparison, since a vanishing quadratic principal symbol about the Levi-Civita locus need not determine the full nonlinear constraint structure \cite{Banados:2025sww,Banados:2025uhb}. Scalar-coupled Palatini Gauss--Bonnet models have also been studied through cosmological solutions and order reduction \cite{Hassan:2026meg}. The present calculation is performed before any such reduction and isolates the connection-level two-derivative structure of the unreduced affine theory.

The phrase ``Palatini Gauss--Bonnet'' therefore does not by itself specify a unique affine theory; the continuation of the Riemannian Gauss--Bonnet combination must be stated. This matters for the local metric scalar--tensor reduction of \cite{BouzariNezhad:2025bgx}, which relies on purely algebraic elimination of the independent connection. The nonzero metric-contracted principal symbol obstructs that elimination in the directions activated by the witness calculation, while the double-epsilon continuation has no corresponding quadratic two-derivative obstruction about $L^{\gamma}{}_{\alpha\beta}=0$.

Section \ref{Sec: Principal-null directions} examines the six directions that the metric-contracted form leaves null in the hook-free timelike restriction. On the five spin-2 directions, the Einstein--Hilbert algebraic form is nonsingular and positive definite for $M^{2}>0$. On the scalar direction, the Einstein--Hilbert self-coupling vanishes while an algebraic cross-coupling remains, so in the null-adapted basis the corresponding variable is Lagrange-multiplier-like for the two quadratic forms evaluated here. Lower-order terms, both those proportional to $\partial_{\alpha}f$ and the zero-derivative Gauss--Bonnet terms built with the background Levi-Civita curvature, can enter the equations along these directions without changing the two-derivative principal symbol, while the omitted hook variables may couple to them already at principal order. These unresolved features do not alter the nonzero two-derivative principal form already established on the activated directions of the hook-free witness subspace.

Several extensions lie outside this calculation. The block survey was carried out for timelike principal covectors; spacelike and null covectors remain to be examined. Restoring the hook sector extends the analysis within the torsion-free connection space, while allowing torsion is a separate enlargement of the theory. This analysis is not a Dirac--Hamiltonian analysis and does not determine the number of propagating degrees of freedom associated with the independent connection. It shows that two affine continuations that coincide in Riemannian geometry already lead to different quadratic two-derivative structures in the connection sector about the Levi-Civita locus.
\appendix
\section{Component evaluation of the quadratic forms}
\label{AppSec: Component evaluation of the quadratic forms}
This appendix gives explicit component realizations of three results used in the main text: the spin-1 and spin-0 blocks of the metric-contracted principal form, the double-epsilon cancellation in the twenty-four-component hook-free parametrization, and the Einstein--Hilbert algebraic matrix on the five spin-2 principal-null directions. These are component checks of results derived and interpreted in the main text.
\subsection{Spin-1 and spin-0 component reconstruction}
\label{AppSubsec: Spin-1 and spin-0 component reconstruction}
The spin-1 and spin-0 blocks are the two mixed blocks of the metric-contracted principal form, and we reconstruct them here from the covariant expression. Substituting the rest-frame decomposition \eqref{Eq: Timelike decomposition of Omega} into the covariant hook-free form \eqref{Eq: Exact hook-free principal quadratic form} and setting all variables outside the spin-1 sector to zero gives, for one fixed spatial direction $i$ with no sum on that index,
\begin{eqnarray}
\label{Eq: Appendix spin-1 component form}
\frac{\widehat{\mathcal{G}}^{(1)}_{\rm{met}}}{\omega^{2}}=\frac{13}{54}P_{i}^{2}+\frac{23}{9}P_{i}\widetilde{Q}_{i}+6\widetilde{Q}_{i}^{2}+P_{i}v_{i}+4\widetilde{Q}_{i}v_{i}+\frac{6}{5}v_{i}^{2}.
\end{eqnarray}

With $\mathbf{x}^{(1)}_{i}=(v_{i},P_{i},\widetilde{Q}_{i})^{T}$, this polynomial is the expansion of $\mathbf{x}^{(1)T}_{i}M^{(1)}\mathbf{x}^{(1)}_{i}$ with the matrix $M^{(1)}$ of \eqref{Eq: Spin-1 block matrix}.

The same substitution in the spin-0 sector gives
\begin{eqnarray}
\label{Eq: Appendix spin-0 component form}
\frac{\widehat{\mathcal{G}}^{(0)}_{\rm{met}}}{\omega^{2}}=\frac{7}{36}P_{0}^{2}+\frac{2}{3}P_{0}\widetilde{Q}_{0}-P_{0}s-4\widetilde{Q}_{0}s-s^{2}.
\end{eqnarray}

With $\mathbf{x}^{(0)}=(s,P_{0},\widetilde{Q}_{0})^{T}$, this is the expansion of $\mathbf{x}^{(0)T}M^{(0)}\mathbf{x}^{(0)}$ with $M^{(0)}$ of \eqref{Eq: Spin-0 block matrix}. The component polynomials therefore reproduce both matrices used in the timelike block analysis of Sec. \ref{Sec: Timelike block decomposition}.
\subsection{Double-epsilon cancellation in the hook-free parametrization}
\label{AppSubsec: Double-epsilon cancellation in the hook-free parametrization}
The vanishing of the double-epsilon principal form was proved analytically in Sec. \ref{Sec: Double-epsilon representative} for an arbitrary distortion perturbation about $L^{\gamma}{}_{\alpha\beta}=0$, without the hook-free restriction, a timelike restriction on the principal covector, or torsion-free symmetry of the perturbation. The metric-contracted blocks above were evaluated in the twenty-four-component hook-free parametrization, and evaluating the double-epsilon form in the same variables gives a direct component-level comparison of the two representatives.

The hook-free sector contains twenty-four independent amplitudes: sixteen in $\Omega_{\alpha\beta\gamma}$, four in $P_{\alpha}$, and four in $\widetilde{Q}_{\alpha}$. Expanding $\widehat{\mathcal{E}}_{\rm{Euler}}$ directly in these amplitudes, without decomposing them into spin blocks, gives zero for every quadratic coefficient, in agreement with \eqref{Eq: Vanishing Euler principal symbol}.

This is a component verification of the analytic momentum argument, not an independent proof. The general vanishing follows from \eqref{Eq: Momentum argument}; the calculation here shows how the same cancellation appears in the hook-free variables used for the metric-contracted principal form.
\subsection{Einstein--Hilbert spin-2 block}
\label{AppSubsec: Einstein--Hilbert spin-2 block}
The main text uses the eigenvalues of the Einstein--Hilbert algebraic form on the five spin-2 principal-null directions; the underlying matrix is given here. In the independent coordinates $\mathbf{x}^{(2)}=(S_{11},S_{22},S_{12},S_{13},S_{23})^{T}$, with $S_{33}=-S_{11}-S_{22}$, the Einstein--Hilbert form \eqref{Eq: Einstein-Hilbert algebraic form} is $\mathcal{B}^{(2)}_{\rm{EH}}=\mathbf{x}^{(2)T}H^{(2)}\mathbf{x}^{(2)}$, with
\begin{eqnarray}
\label{Eq: Appendix Einstein-Hilbert spin-2 matrix}
H^{(2)}=\frac{M^{2}}{2}
\begin{pmatrix}
\dfrac{3}{2} & \dfrac{3}{4} & 0 & 0 & 0\\[6pt]
\dfrac{3}{4} & \dfrac{3}{2} & 0 & 0 & 0\\[6pt]
0 & 0 & \dfrac{3}{2} & 0 & 0\\[6pt]
0 & 0 & 0 & \dfrac{3}{2} & 0\\[6pt]
0 & 0 & 0 & 0 & \dfrac{3}{2}
\end{pmatrix}.
\end{eqnarray}

Its determinant is $(M^{2}/2)^{5}729/128$, which is nonzero, and diagonalizing it gives the eigenvalues quoted in \eqref{Eq: Einstein-Hilbert spin-2 eigenvalues}, all positive for $M^{2}>0$. This is the explicit component check that the Einstein--Hilbert algebraic form is nonsingular and positive definite on these five directions.
\bibliographystyle{apsrev4-2}
\bibliography{References}

@article{Lanczos:1938sf,
	author = "Lanczos, Cornelius",
	title = "{A Remarkable property of the Riemann-Christoffel tensor in four dimensions}",
	doi = "10.2307/1968467",
	journal = "Annals Math.",
	volume = "39",
	pages = "842--850",
	year = "1938"
}

@article{Lovelock:1971yv,
	author = "Lovelock, D.",
	title = "{The Einstein tensor and its generalizations}",
	doi = "10.1063/1.1665613",
	journal = "J. Math. Phys.",
	volume = "12",
	pages = "498--501",
	year = "1971"
}

@article{Deruelle:2018vtt,
	author = "Deruelle, Nathalie and Merino, Nelson and Olea, Rodrigo",
	title = "{Chern-Weil theorem, Lovelock Lagrangians in critical dimensions and boundary terms in gravity actions}",
	eprint = "1803.04741",
	archivePrefix = "arXiv",
	primaryClass = "gr-qc",
	doi = "10.1103/PhysRevD.98.044031",
	journal = "Phys. Rev. D",
	volume = "98",
	number = "4",
	pages = "044031",
	year = "2018"
}

@article{Horndeski:1974wa,
	author = "Horndeski, Gregory Walter",
	title = "{Second-order scalar-tensor field equations in a four-dimensional space}",
	doi = "10.1007/BF01807638",
	journal = "Int. J. Theor. Phys.",
	volume = "10",
	pages = "363--384",
	year = "1974"
}

@article{Kobayashi:2011nu,
	author = "Kobayashi, Tsutomu and Yamaguchi, Masahide and Yokoyama, Jun'ichi",
	title = "{Generalized G-inflation: Inflation with the most general second-order field equations}",
	eprint = "1105.5723",
	archivePrefix = "arXiv",
	primaryClass = "hep-th",
	reportNumber = "KUNS-2339, RESCEU-9-11",
	doi = "10.1143/PTP.126.511",
	journal = "Prog. Theor. Phys.",
	volume = "126",
	pages = "511--529",
	year = "2011"
}

@article{Borunda:2008kf,
	author = "Borunda, Monica and Janssen, Bert and Bastero-Gil, Mar",
	title = "{Palatini versus metric formulation in higher curvature gravity}",
	eprint = "0804.4440",
	archivePrefix = "arXiv",
	primaryClass = "hep-th",
	reportNumber = "UG-FT-225-08, CAFPE-95-08",
	doi = "10.1088/1475-7516/2008/11/008",
	journal = "JCAP",
	volume = "11",
	pages = "008",
	year = "2008"
}

@article{BeltranJimenez:2014iie,
	author = "Beltran Jimenez, Jose and Koivisto, Tomi S.",
	title = "{Extended Gauss-Bonnet gravities in Weyl geometry}",
	eprint = "1402.1846",
	archivePrefix = "arXiv",
	primaryClass = "gr-qc",
	reportNumber = "NORDITA-2014-14",
	doi = "10.1088/0264-9381/31/13/135002",
	journal = "Class. Quant. Grav.",
	volume = "31",
	pages = "135002",
	year = "2014"
}

@article{Vitagliano:2010sr,
	author = "Vitagliano, Vincenzo and Sotiriou, Thomas P. and Liberati, Stefano",
	title = "{The dynamics of metric-affine gravity}",
	eprint = "1008.0171",
	archivePrefix = "arXiv",
	primaryClass = "gr-qc",
	doi = "10.1016/j.aop.2011.02.008",
	journal = "Annals Phys.",
	volume = "326",
	pages = "1259--1273",
	year = "2011",
	note = "[Erratum: Annals Phys. 329, 186--187 (2013)]"
}

@article{Bauer:2008zj,
	author = "Bauer, Florian and Demir, Durmus A.",
	title = "{Inflation with Non-Minimal Coupling: Metric versus Palatini Formulations}",
	eprint = "0803.2664",
	archivePrefix = "arXiv",
	primaryClass = "hep-ph",
	reportNumber = "DESY-08-033, IZTECH-P-08-02",
	doi = "10.1016/j.physletb.2008.06.014",
	journal = "Phys. Lett. B",
	volume = "665",
	pages = "222--226",
	year = "2008"
}

@article{Rasanen:2017ivk,
	author = "Rasanen, Syksy and Wahlman, Pyry",
	title = "{Higgs inflation with loop corrections in the Palatini formulation}",
	eprint = "1709.07853",
	archivePrefix = "arXiv",
	primaryClass = "astro-ph.CO",
	reportNumber = "HIP-2017-23-TH, KOBE-COSMO-17-12",
	doi = "10.1088/1475-7516/2017/11/047",
	journal = "JCAP",
	volume = "11",
	pages = "047",
	year = "2017"
}

@article{Rasanen:2018ihz,
	author = "Rasanen, Syksy",
	title = "{Higgs inflation in the Palatini formulation with kinetic terms for the metric}",
	eprint = "1811.09514",
	archivePrefix = "arXiv",
	primaryClass = "gr-qc",
	reportNumber = "HIP-2018-27/TH",
	doi = "10.21105/astro.1811.09514",
	journal = "Open J. Astrophys.",
	volume = "2",
	number = "1",
	pages = "1",
	year = "2019"
}

@article{Nezhad:2023dys,
	author = "Nezhad, Hamed Bouzari and Rasanen, Syksy",
	title = "{Scalar fields with derivative coupling to curvature in the Palatini and the metric formulation}",
	eprint = "2307.04618",
	archivePrefix = "arXiv",
	primaryClass = "gr-qc",
	reportNumber = "HIP-2023-11/TH",
	doi = "10.1088/1475-7516/2024/02/009",
	journal = "JCAP",
	volume = "02",
	pages = "009",
	year = "2024"
}

@article{Janssen:2019uao,
	author = "Janssen, Bert and Jim{\'e}nez-Cano, Alejandro",
	title = "{On the topological character of metric-affine Lovelock Lagrangians in critical dimensions}",
	eprint = "1907.12100",
	archivePrefix = "arXiv",
	primaryClass = "gr-qc",
	doi = "10.1016/j.physletb.2019.134996",
	journal = "Phys. Lett. B",
	volume = "798",
	pages = "134996",
	year = "2019"
}

@phdthesis{JimenezCano:2021rlu,
	author = "Jim{\'e}nez Cano, Alejandro",
	title = "{Metric-affine Gauge theories of gravity. Foundations and new insights}",
	eprint = "2201.12847",
	archivePrefix = "arXiv",
	primaryClass = "gr-qc",
	school = "Granada U., Theor. Phys. Astrophys.",
	year = "2021"
}

@article{Iosifidis:2021crj,
	author = "Iosifidis, Damianos",
	title = "{Riemann tensor and Gauss{\textendash}Bonnet density in metric-affine cosmology}",
	eprint = "2104.10192",
	archivePrefix = "arXiv",
	primaryClass = "gr-qc",
	doi = "10.1088/1361-6382/ac213a",
	journal = "Class. Quant. Grav.",
	volume = "38",
	number = "19",
	pages = "195028",
	year = "2021"
}

@article{Hassan:2026meg,
	author = "Hassan, Ali and Rasanen, Syksy",
	title = "{Inflation with the Gauss-Bonnet term in the Palatini formulation}",
	eprint = "2603.17742",
	archivePrefix = "arXiv",
	primaryClass = "astro-ph.CO",
	reportNumber = "HIP-2026-5/TH",
	doi = "10.1088/1475-7516/2026/05/088",
	journal = "JCAP",
	volume = "05",
	pages = "088",
	year = "2026"
}

@article{BeltranJimenez:2019acz,
	author = "Beltr{\'a}n Jim{\'e}nez, Jose and Delhom, Adria",
	title = "{Ghosts in metric-affine higher order curvature gravity}",
	eprint = "1901.08988",
	archivePrefix = "arXiv",
	primaryClass = "gr-qc",
	doi = "10.1140/epjc/s10052-019-7149-x",
	journal = "Eur. Phys. J. C",
	volume = "79",
	number = "8",
	pages = "656",
	year = "2019"
}

@article{BeltranJimenez:2020sqf,
	author = "Beltr{\'a}n Jim{\'e}nez, Jose and Delhom, Adri{\`a}",
	title = "{Instabilities in metric-affine theories of gravity with higher order curvature terms}",
	eprint = "2004.11357",
	archivePrefix = "arXiv",
	primaryClass = "gr-qc",
	doi = "10.1140/epjc/s10052-020-8143-z",
	journal = "Eur. Phys. J. C",
	volume = "80",
	number = "6",
	pages = "585",
	year = "2020"
}

@article{Percacci:2020ddy,
	author = "Percacci, R. and Sezgin, E.",
	title = "{New class of ghost- and tachyon-free metric affine gravities}",
	eprint = "1912.01023",
	archivePrefix = "arXiv",
	primaryClass = "hep-th",
	reportNumber = "MI-TH-1941",
	doi = "10.1103/PhysRevD.101.084040",
	journal = "Phys. Rev. D",
	volume = "101",
	number = "8",
	pages = "084040",
	year = "2020",
	note = "[Erratum: Phys.Rev.D 111, 109902 (2025)]"
}

@article{Hehl:1994ue,
	author = "Hehl, Friedrich W. and McCrea, J. Dermott and Mielke, Eckehard W. and Ne'eman, Yuval",
	title = "{Metric affine gauge theory of gravity: Field equations, Noether identities, world spinors, and breaking of dilation invariance}",
	eprint = "gr-qc/9402012",
	archivePrefix = "arXiv",
	reportNumber = "TAUP-N192-94, TAUP-192-94",
	doi = "10.1016/0370-1573(94)00111-F",
	journal = "Phys. Rept.",
	volume = "258",
	pages = "1--171",
	year = "1995"
}

@article{Shimada:2018lnm,
	author = "Shimada, Keigo and Aoki, Katsuki and Maeda, Kei-ichi",
	title = "{Metric-affine Gravity and Inflation}",
	eprint = "1812.03420",
	archivePrefix = "arXiv",
	primaryClass = "gr-qc",
	reportNumber = "WU-AP/1808/18",
	doi = "10.1103/PhysRevD.99.104020",
	journal = "Phys. Rev. D",
	volume = "99",
	number = "10",
	pages = "104020",
	year = "2019"
}

@article{Reall:2014pwa,
	author = "Reall, Harvey and Tanahashi, Norihiro and Way, Benson",
	title = "{Causality and Hyperbolicity of Lovelock Theories}",
	eprint = "1406.3379",
	archivePrefix = "arXiv",
	primaryClass = "hep-th",
	doi = "10.1088/0264-9381/31/20/205005",
	journal = "Class. Quant. Grav.",
	volume = "31",
	pages = "205005",
	year = "2014"
}

@article{Papallo:2017qvl,
	author = "Papallo, Giuseppe and Reall, Harvey S.",
	title = "{On the local well-posedness of Lovelock and Horndeski theories}",
	eprint = "1705.04370",
	archivePrefix = "arXiv",
	primaryClass = "gr-qc",
	doi = "10.1103/PhysRevD.96.044019",
	journal = "Phys. Rev. D",
	volume = "96",
	number = "4",
	pages = "044019",
	year = "2017"
}

@article{McCrea:1992wa,
	author = "McCrea, J. D.",
	title = "{Irreducible decompositions of non-metricity, torsion, curvature and Bianchi identities in metric affine space-times}",
	doi = "10.1088/0264-9381/9/2/018",
	journal = "Class. Quant. Grav.",
	volume = "9",
	pages = "553--568",
	year = "1992"
}

@article{Paci:2023twc,
	author = "Paci, Gregorio and Sauro, Dario and Zanusso, Omar",
	title = "{Conformally covariant operators of mixed-symmetry tensors and MAGs}",
	eprint = "2302.14093",
	archivePrefix = "arXiv",
	primaryClass = "hep-th",
	doi = "10.1088/1361-6382/acf9d8",
	journal = "Class. Quant. Grav.",
	volume = "40",
	number = "21",
	pages = "215005",
	year = "2023"
}

@article{Mikura:2024mji,
	author = "Mikura, Yusuke and Percacci, Roberto",
	title = "{Some simple theories of gravity with propagating nonmetricity}",
	eprint = "2401.10097",
	archivePrefix = "arXiv",
	primaryClass = "gr-qc",
	doi = "10.1140/epjc/s10052-025-14036-w",
	journal = "Eur. Phys. J. C",
	volume = "85",
	number = "4",
	pages = "377",
	year = "2025"
}

@article{Janssen:2019doc,
	author = "Janssen, Bert and Jim{\'e}nez-Cano, Alejandro and Orejuela, Jos{\'e} Alberto",
	title = "{A non-trivial connection for the metric-affine Gauss-Bonnet theory in $D = 4$}",
	eprint = "1903.00280",
	archivePrefix = "arXiv",
	primaryClass = "gr-qc",
	doi = "10.1016/j.physletb.2019.06.002",
	journal = "Phys. Lett. B",
	volume = "795",
	pages = "42--48",
	year = "2019"
}

@article{Baldazzi:2021kaf,
	author = "Baldazzi, A. and Melichev, O. and Percacci, R.",
	title = "{Metric-Affine Gravity as an effective field theory}",
	eprint = "2112.10193",
	archivePrefix = "arXiv",
	primaryClass = "gr-qc",
	doi = "10.1016/j.aop.2022.168757",
	journal = "Annals Phys.",
	volume = "438",
	pages = "168757",
	year = "2022"
}

@article{Baekler:2006vw,
	author = "Baekler, Peter and Boulanger, Nicolas and Hehl, Friedrich W.",
	title = "{Linear connections with propagating spin-3 field in gravity}",
	eprint = "hep-th/0608122",
	archivePrefix = "arXiv",
	doi = "10.1103/PhysRevD.74.125009",
	journal = "Phys. Rev. D",
	volume = "74",
	pages = "125009",
	year = "2006"
}

@article{Percacci:2025oxw,
	author = "Percacci, Roberto and Sezgin, Ergin",
	title = "{Massive spin 3 and Metric-Affine Gravity}",
	eprint = "2508.14211",
	archivePrefix = "arXiv",
	primaryClass = "hep-th",
	reportNumber = "MI-HET-863",
	doi = "10.1007/JHEP01(2026)042",
	journal = "JHEP",
	volume = "01",
	pages = "042",
	year = "2026"
}

@article{Glavan:2023cuy,
	author = "Glavan, Dra{\v{z}}en and Zlosnik, Tom and Lin, Chunshan",
	title = "{Hamiltonian analysis of metric-affine-$R^{2}$ theory}",
	eprint = "2311.17459",
	archivePrefix = "arXiv",
	primaryClass = "gr-qc",
	doi = "10.1088/1475-7516/2024/04/072",
	journal = "JCAP",
	volume = "04",
	pages = "072",
	year = "2024"
}

@article{Banados:2025sww,
	author = "Banados, Maximo and Henneaux, Marc",
	title = "{Palatini Gauss-Bonnet theory}",
	eprint = "2511.20903",
	archivePrefix = "arXiv",
	primaryClass = "hep-th",
	doi = "10.1007/JHEP03(2026)064",
	journal = "JHEP",
	volume = "03",
	pages = "064",
	year = "2026"
}

@article{Banados:2025uhb,
	author = "Ba{\~n}ados, M{\'a}ximo and Bennett, Daniela",
	title = "{Spherical symmetric fields on torsion-free Palatini Gauss-Bonnet theory}",
	eprint = "2512.12863",
	archivePrefix = "arXiv",
	primaryClass = "gr-qc",
	doi = "10.1088/1361-6382/ae5164",
	journal = "Class. Quant. Grav.",
	volume = "43",
	pages = "065015",
	year = "2026"
}

@article{BouzariNezhad:2025bgx,
	author = "Bouzari Nezhad, Hamed",
	title = "{Degenerate higher-order scalar-tensor theories in metric-affine gravity}",
	eprint = "2512.08972",
	archivePrefix = "arXiv",
	primaryClass = "gr-qc",
	doi = "10.1088/1475-7516/2026/04/056",
	journal = "JCAP",
	volume = "04",
	pages = "056",
	year = "2026"
}
\end{document}